\documentclass[%
reprint,
jmp,
superscriptaddress,
nofootinbib,
amsmath,amssymb,
table
]{revtex4-2}
\usepackage[english]{babel}
\usepackage{graphicx}
\usepackage{dcolumn}
\usepackage{bm}
\usepackage[version=4]{mhchem}
\usepackage{svg}
\usepackage{multirow}
\usepackage{comment}
\usepackage{tabularx}
\usepackage{array}
\usepackage{makecell}
\usepackage{amssymb}
\usepackage{float}
\usepackage{afterpage}
\usepackage[utf8]{inputenc}
\usepackage{braket}
\usepackage{amsmath}
\usepackage{amsfonts}
\usepackage{booktabs} 
\usepackage{titlesec}
\usepackage{dsfont}
\usepackage{pifont}
\usepackage{adjustbox}
\usepackage[normalem]{ulem}
\usepackage{tabularray}
\usepackage[left=0.8in,right=0.8in,top=0.8in,bottom=0.8in]{geometry}
\usepackage{upgreek}
\usepackage{textgreek}
\usepackage[super]{nth}
\usepackage{csquotes}
\usepackage{xcolor}
\usepackage{xr}
\usepackage{appendix}
\usepackage{listings}
\usepackage{hhline}
\usepackage{hyperref}
\definecolor{codegreen}{rgb}{0,0.6,0}
\definecolor{codegray}{rgb}{0.5,0.5,0.5}
\definecolor{codepurple}{rgb}{0.58,0,0.82}
\definecolor{backcolour}{rgb}{0.95,0.95,0.92}

\lstdefinestyle{bashstyle}{
    backgroundcolor=\color{backcolour},   
    commentstyle=\color{codegreen},
    keywordstyle=\color{magenta},
    numberstyle=\tiny\color{codegray},
    stringstyle=\color{codepurple},
    basicstyle=\ttfamily\footnotesize,
    breakatwhitespace=false,         
    breaklines=true,                 
    captionpos=b,                    
    keepspaces=true,                 
    numbers=left,                    
    numbersep=5pt,                  
    showspaces=false,                
    showstringspaces=false,
    showtabs=false,                  
    tabsize=2,
    frame=single,
    rulecolor=\color{black}
}

\newcolumntype{Y}{>{\centering\arraybackslash}X}

\definecolor{myblue}{RGB}{71,120,207}
\definecolor{mygreen}{RGB}{106,204,100}
\definecolor{myred}{RGB}{213,95,95}

\date{\today}

\makeatletter
\def\@email#1#2{%
 \endgroup
 \patchcmd{\titleblock@produce}
  {\frontmatter@RRAPformat}
  {\frontmatter@RRAPformat{\produce@RRAP{*#1\href{mailto:#2}{#2}}}\frontmatter@RRAPformat}
  {}{}
}%
\makeatother

\begin{document}

\preprint{AIP/123-QED}

\title{DFT Accuracy on Crystal Structure Prediction with Machine Learning Interatomic Potentials}
\author{Laurence I. Midgley}
\affiliation{Ångström AI, San Francisco, USA}
\affiliation{Engineering Laboratory, University of Cambridge, Trumpington Street, CB2 1PZ, Cambridge, UK}
\author{Chen Lin}
\affiliation{Ångström AI, San Francisco, USA}
\affiliation{Engineering Laboratory, University of Cambridge, Trumpington Street, CB2 1PZ, Cambridge, UK}
\author{J. Harry Moore}
\affiliation{Ångström AI, San Francisco, USA}
\affiliation{Engineering Laboratory, University of Cambridge, Trumpington Street, CB2 1PZ, Cambridge, UK}
\author{Flaviano Della Pia}
\affiliation{Ångström AI, San Francisco, USA}
\affiliation{Engineering Laboratory, University of Cambridge, Trumpington Street, CB2 1PZ, Cambridge, UK}
\author{Javier Antor\'an}
\affiliation{Ångström AI, San Francisco, USA}
\affiliation{Engineering Laboratory, University of Cambridge, Trumpington Street, CB2 1PZ, Cambridge, UK}
\author{Sten O. Nilsson Lill}
\affiliation{Predictive Science, Digital and Automation, Pharmaceutical Sciences R$\&$D, AstraZeneca, Gothenburg, Sweden}
\author{Emma S. E. Eriksson}
\affiliation{Predictive Science, Digital and Automation, Pharmaceutical Sciences R$\&$D, AstraZeneca, Gothenburg, Sweden}
\author{Felix A. Faber}
\affiliation{Predictive Science, Digital and Automation, Pharmaceutical Sciences R$\&$D, AstraZeneca, Gothenburg, Sweden}
\author{Lars Tornberg}
\affiliation{Predictive Science, Digital and Automation, Pharmaceutical Sciences R$\&$D, AstraZeneca, Gothenburg, Sweden}
\author{Anders Broo}
\affiliation{Predictive Science, Digital and Automation, Pharmaceutical Sciences R$\&$D, AstraZeneca, Gothenburg, Sweden}
\author{G\'abor Cs\'anyi}
\affiliation{Ångström AI, San Francisco, USA}
\affiliation{Max Planck Institute for Polymer Research, Ackermannweg 10, Mainz, 55128, Germany}
\affiliation{Engineering Laboratory, University of Cambridge, Trumpington Street, CB2 1PZ, Cambridge, UK}

\begin{abstract}
\section*{Abstract}

We present an evaluation of CSP-MACE-Å, a machine learning interatomic potential intended to replace DFT in crystal structure prediction (CSP). 
We decompose the total energy into separate intramolecular and intermolecular components, allowing each component to be designed appropriately for its interaction type.
For the intramolecular component, we adopt the MACE-POLAR architecture and train it on the OMol25 dataset.
The intermolecular model is designed to capture the subtle intermolecular interactions in crystal structures. 
This intermolecular component combines three terms: an intermolecular contribution from the MACE-POLAR model, a long-range dispersion term with the functional form of the XDM correction, and a learned delta model trained to reproduce B86bPBE-XDM intermolecular energies.
The learned delta model is trained on residual intermolecular targets derived from 50,000 B86bPBE-XDM calculations on molecular crystal structures.
On an evaluation set composed of 19 compounds, including a salt, selected from AstraZeneca's previous CSP publications, CSP-MACE-Å achieves performance comparable to PBE DFT with the Neumann–Perrin dispersion correction.
On a second evaluation set composed of 28 compounds, including cocrystals and salts, collated from the seven CSP blind tests, CSP-MACE-Å achieves performance close to B86bPBE-XDM DFT. 
In both evaluation sets, reranking with harmonic free energies substantially improves performance relative to ranking by energy alone. 
Across our evaluation suite, CSP-MACE-Å is shown to outperform the MACE-POLAR-1 and UMA-OMC foundation models. 
Lastly, on a set of five compounds, CSP-MACE-Å is shown to capture temperature-dependent trends in the relative stability of polymorphs through estimation of the free energy under the harmonic approximation. 
By running multiple orders of magnitude faster than DFT, CSP-MACE-Å enables energy and free energy evaluation of far more candidate structures, providing greater confidence when derisking solid forms.

\end{abstract}

\maketitle

\section{Introduction}


Solid-state form selection of an active pharmaceutical ingredient (API) is a critical aspect of drug development, as it affects bioavailability, manufacturability, and stability. 
Many drug- molecules exhibit crystal polymorphism, where different crystal forms possess distinct physicochemical properties such as solubility and dissolution behavior, making form selection both complex and essential.
Polymorphism presents notable risks, particularly through late-appearing forms that may emerge during manufacturing or storage and alter product performance. 
Such events can lead to significant development and regulatory challenges, underscoring the importance of understanding the polymorphic landscape.
Although experimental screening is the primary method for identifying suitable polymorphs, it may not capture all relevant forms. 
Computational approaches, especially crystal structure prediction (CSP), provide a valuable complement by enabling a more comprehensive exploration of possible crystal structures \cite{broo2016transferable}. 
Together, these strategies support robust form selection and reduce the risk of unexpected solid-state transformations.
\\

CSP is generally broken down into two phases: structure generation and energy ranking. 
During generation, a large quantity of candidate structures (often in the millions) are produced. 
This is typically done by enumerating possible space groups and the number of formula units per cell in combination with random structure generation, to cover the crystal structure landscape. 
A preliminary ranking of the structures is performed via a computationally inexpensive method (e.g. a classical force field) and a small fraction of these, on the order of 1000 structures, are then passed onto the ranking phase. 
During the ranking phase, the structures are ranked using more computationally expensive and accurate techniques, to determine which are the most stable.
The state-of-the-art for this phase is dispersion-corrected DFT (DFT-D), which has consistently dominated the seven CSP blind tests \cite{lommerse2000test,motherwell2002crystal,day2005third,day2009significant,bardwell2011towards,reilly2016report,hunnisett2024seventh_rank,hunnisett2024seventh_gen}.
Further accuracy at greater computational cost is often obtained by reranking with thermal free energies instead of lattice energies \cite{firaha2023predicting}. 
\\

Despite its accuracy, DFT remains computationally expensive and slow. 
Crystal structure optimization typically takes hours, and free energy calculations can take days. 
Machine learning interatomic potentials (MLIPs) offer the possibility to replace DFT in CSP, potentially accelerating calculations by four orders of magnitude, while achieving close to DFT level accuracy. 
The most recent blind test was the first to include MLIPs; however, they showed subpar performance relative to DFT \cite{hunnisett2024seventh_rank}. 
The key challenges for MLIPs in CSP are the need to model long-range electrostatics and dispersion, and to accurately capture subtle intermolecular interactions. 
The recently released MACE-POLAR-1 model is an MLIP that includes a principled treatment of long-range electrostatic interactions \cite{batatia2026macepolar}. 
However, this model does not describe the long-range tail of dispersion, and is trained on the OMol25 dataset \cite{levine2025open} which does not include crystal structures. 
In this work, we present an evaluation of CSP-MACE-Å which addresses these challenges to current MLIPs in CSP. 
\\

The promise of a fast surrogate is not only to replace DFT in existing CSP workflows, but also to change the workflows themselves.
The increased speed will allow more candidate structures to be studied in the ranking stage of CSP.
This could improve accuracy in cases where viable polymorphs were excluded from the ranking stage due to the prohibitive cost of DFT \cite{nickerson2025assessment}. 
Similarly, it could make free energy reranking practical in settings where the cost of DFT prohibits many structures (or any structures) from having their free energy computed. 
More broadly, the speed of an ML surrogate enables its use earlier in the drug development pipeline, and pairing it with property prediction (e.g. intrinsic solubility) would further extend its utility.

\section{Methods}

\subsection{CSP-MACE-Å}
The predicted energy of CSP-MACE-Å is decomposed into separate intermolecular and intramolecular components:
\begin{equation}
 E_{total}^{\text{CSP-MACE-Å}} = E_{intra}^{\text{CSP-MACE-Å}} + E_{inter}^{\text{CSP-MACE-Å}}
\end{equation}
which provides flexibility to design each component of the model appropriately for the interaction type. 
We define the intramolecular energy as the sum of the energies of the constituent molecules in a crystal structure each isolated in vacuum. 
We then define the intermolecular energy as: 
\begin{equation}
\label{eqn:inter_intra}
 E_{inter} = E_{total} - E_{intra}
\end{equation}
where $E_{total}$ is the energy of the full system with periodic boundary conditions 
and $E_{intra}$ is the intramolecular energy. 
This definition is practical rather than canonical and allows us to easily decompose any energy function into its intramolecular and intermolecular contributions. 
This type of decomposition is common in the CSP literature and is sometimes referred to as \textit{monomer conformational energy corrections}. 
It is typically used to combine higher level theory DFT or post-Hartree–Fock methods for the intramolecular term with less expensive DFT methods for the intermolecular term \cite{greenwell2020inaccurate,hunnisett2024seventh_rank}.

\subsection{Intramolecular model}
For the intramolecular component of CSP-MACE-Å, we train the MACE-POLAR architecture described in \cite{batatia2026macepolar}, with the medium size setting, on the OMol25 dataset \cite{levine2025open}. 
The MACE-POLAR architecture builds upon the local MACE architecture to include principled treatment of long-range electrostatics, which is critical for CSP. 
The OMol25 dataset is composed of 100 million ωB97M-V/def2-TZVPD \cite{mardirossian2016omegab97m,rappoport2010property} DFT calculations covering a wide range of chemical space. 
The high level of theory and scale of the dataset make it well suited for training a model on intramolecular interactions. 
\\

However, the MACE-POLAR architecture by itself is insufficient as it does not include long-range dispersion, which is of critical importance to CSP. 
Furthermore, the OMol25 dataset on which the MACE-POLAR model is trained is a poor choice for intermolecular modelling in CSP, as it does not contain any crystal structures.
In addition, OMol25 was generated using ωB97M-V, whose VV-10 dispersion is known to exhibit overbinding \cite{hujo2011performance}; the MACE-POLAR model therefore inherits this bias. 

\subsection{Intermolecular model}
For the intermolecular component of CSP-MACE-Å, we target B86bPBE-XDM DFT \cite{becke1986large,perdew1996generalized,becke2007exchange,johnson2017exchange} because it has been shown to capture intermolecular interactions in crystals accurately. 
For example, it performs well on the seven CSP blind tests \citep{price2023accurate,mayo2024assessment} 
and achieves state-of-the-art lattice energy accuracy on the X23 dataset relative to a diffusion Monte Carlo reference \cite{della2025accurate}. 
\\

To construct the intermolecular model we combine the intermolecular contribution of the MACE-POLAR model with a dispersion model and a learned delta model:
\begin{equation}
E_{inter}^{\text{CSP-MACE-Å}} = E_{inter}^{\text{MACE-POLAR}} + E_{inter}^{\text{dispersion}} + E_{inter}^{\text{delta-model}}
\end{equation}
where each $E_{inter}$ term itself follows the functional form defined in \autoref{eqn:inter_intra}. 
Our dispersion model follows the functional form of the XDM dispersion correction:
\begin{equation}
E^{\mathrm{XDM}} = -\frac{1}{2}\sum_{n=6,8,10}\sum_{ij}\frac{C_{n,ij}}{R_{ij}^{n} + R_{\mathrm{vdW},ij}^{n}},
\end{equation}
where $R_{ij}$ is the interatomic distance, $C_6$, $C_8$, and $C_{10}$ are the XDM atomic dispersion coefficients and $R_{\mathrm{vdW}}$ is the sum of the effective van der Waals radii of atoms $i$ and $j$. 
For our dispersion model, the parameters $C_{6,ij}$, $C_{8,ij}$, $C_{10,ij}$, $R_{\mathrm{vdW},ij}$ for each atom pair ${\{i, j\}}$ are fixed to average value estimated from Angstrom AI's internal 50,000 set of B86bPBE-XDM DFT calculations.
This provides us with a computationally inexpensive approximation to the XDM-dispersion correction. 
The purpose of the dispersion model is to accurately capture long-range intermolecular interactions. 
We rely on other components of $E_{inter}^{\text{CSP-MACE-Å}}$ (i.e. $E_{inter}^{\text{MACE-POLAR}} + E_{inter}^{\text{delta-model}}$) to capture short-range intermolecular interactions where more expressive power is required from the model. 
\\

The intermolecular delta model, $E_{inter}^{\text{delta-model}}$, corrects the other components of the intermolecular model to match B86bPBE-XDM DFT. 
We therefore train the intermolecular delta model to predict the residual between the B86bPBE-XDM DFT labels and the sum of the MACE-POLAR model and dispersion model contributions.
The energy target for the delta model is
\begin{equation}
E_{inter}^{\text{delta-target}} = E_{inter}^{\text{DFT}} - E_{inter}^{\text{MACE-POLAR}}- E_{inter}^{\text{dispersion}} . 
\end{equation}
Analogous residual targets are defined for the force and stress losses.
We train the intermolecular delta model on Ångstrom AI's internal dataset of 50,000 B86bPBE-XDM DFT crystal calculations. 
Intermolecular contributions to the energy and forces are more than an order of magnitude smaller than intramolecular contributions. 
A key feature of CSP-MACE-Å parameterization and training is that the signal for the intermolecular interactions is isolated from the intramolecular interactions, which would otherwise dominate the loss - as is the case in standard MLIP training on total targets. 

\section{Results and Discussion}

\subsection{AZ Set}
\label{sec:az_eval_set}

\begin{figure*}[t]
    \centering
    \includegraphics[width=\linewidth]{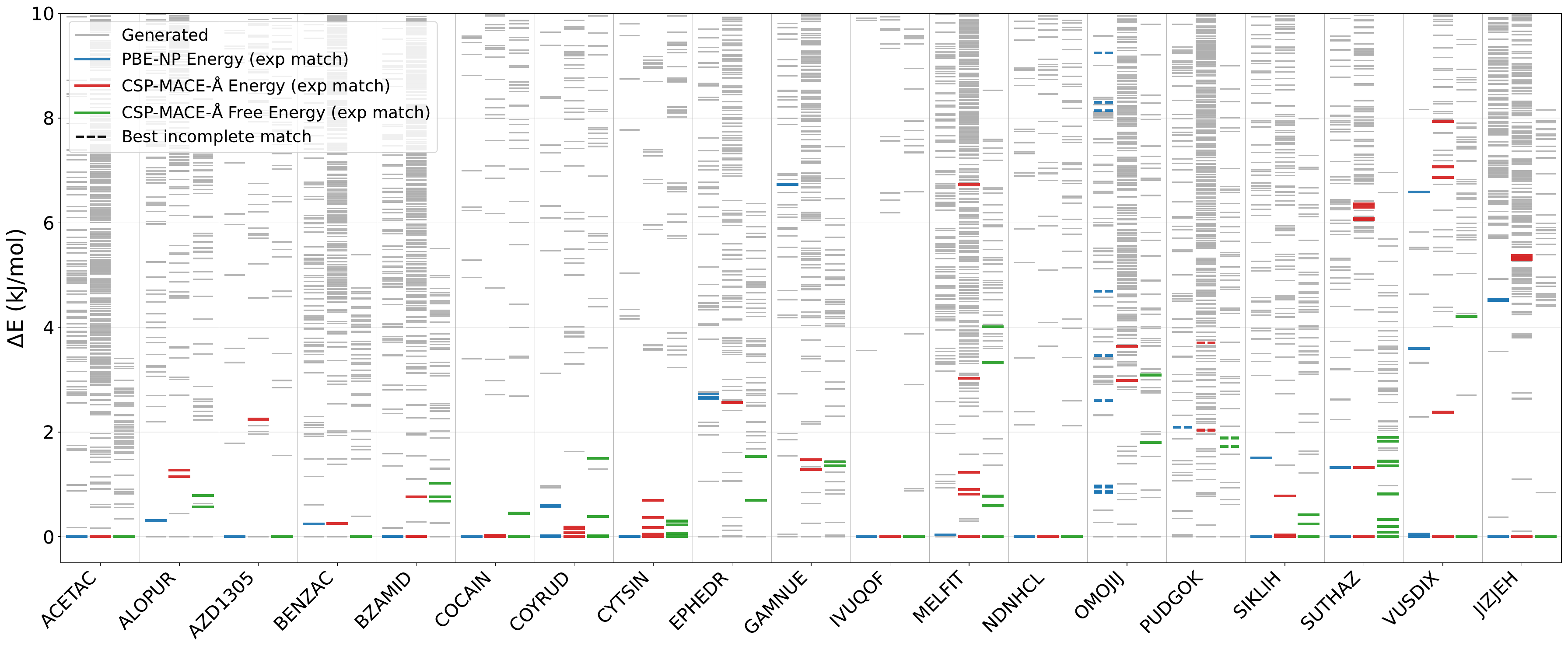}
    \caption{AZ evaluation set relative energy landscape. The left hand column for each compound is PBE-NP DFT Energy ranking and the central column CSP-MACE-Å energy ranking, and the right hand column is CSP-MACE-Å free energy ranking under the harmonic approximation at 300 K. Generated structures which match an experimental polymorph are shown in colour while those without a match are shown in grey. We use COMPACK \cite{chisholm2005compack}, with a setting of $RMSD_{15} \mathord{<} 0.6\text{ \AA}$, to determine which relaxed structures match the experimental structures for each compound. For compounds where there are no experimental matches, the best partial matches are shown where best is defined as matches with the largest cluster size $X$ such that $RMSD_{X} \mathord{<} 0.6\text{ \AA}$. JIZJEH (shown in the final column) is a salt.}
    \label{fig:az_landscape}
\end{figure*}

First, we evaluate CSP-MACE-Å on a set of 19 compounds, including a salt, from a selection of AstraZeneca's CSP publications \cite{broo2016transferable,andrews2021derisking,sun2021current}. 
End-to-end CSP has previously been carried out on this set. 
Millions of candidate structures were generated for each compound using GRACE \cite{neumann2008major} and then optimized and ranked using AstraZeneca's CSP forcefield, AZ-FF \cite{broo2016transferable}. 
The $\sim$100 lowest\footnote{\label{fn:shared} As this set combines CSP from multiple previous publications, in some cases there were less than 1000 AZ-FF relaxed structures available for MLIP reranking, and less than 100 structures on which PBE-NP reranking was performed. Additionally, for some compounds single-point energy calculations based on DFT with the PBE functional and the Tkatchenko–Scheffler dispersion correction \cite{tkatchenko2009accurate} were used for reranking of the AZ-FF structures before the final relaxation and reranking with the PBE-NP. 
We provide further details in \autoref{app:az-eval-set}.} 
energy structures were then re-optimized and reranked using DFT with the PBE functional \cite{perdew1996generalized} and the Neumann–Perrin dispersion correction \cite{ neumann2005energy} which we refer to as PBE-NP. 
We evaluate CSP-MACE-Å by substituting it for DFT, enabling direct comparison with the existing ranking protocol.
Because CSP-MACE-Å is substantially cheaper than DFT, we apply it to a larger pool of candidate structures. 
Specifically, we use CSP-MACE-Å to relax and rerank the $\sim$1000 lowest\footnotemark[\value{footnote}] AZ-FF ranked structures, and compare this to the PBE-NP ranking.
We further benchmark two additional models. 
The first is the MACE-POLAR-1 medium model \cite{batatia2026macepolar}, a key architectural component of CSP-MACE-Å. 
The second is UMA-OMC\footnote{UMA-OMC refers to the UMA-S-1.1 model trained on the OMC25 dataset of over 25 million configurations of molecular crystals evaluated with PBE+D3 DFT.} \cite{gharakhanyan2026open}, which represents the state-of-the-art MLIP for crystal structure prediction \cite{gharakhanyan2025fastcsp}. 
Finally, we include an additional step, whereby we calculate free energies using the MLIPs under the harmonic approximation and perform another stage of reranking. 
\\

We use COMPACK \cite{chisholm2005compack}, with a setting of $RMSD_{15} \mathord{<} 0.6\text{ \AA}$, to determine which relaxed structures match the experimental structures for each compound. 
For polymorphic compounds (MELFIT, SUTHAZ, VUSDIX, JIZJEH, AZD1305, NDNHCL and SIKLIH) multiple experimental structures are included in the matching. 
Further details are provided in appendix \ref{app:az-eval-set}. 
For compounds where there is no experimental match, we report results for partial matches.
To asses the ranking accuracy, we compare the rank of the experimental match and $\Delta E$ defined as the difference in energy between the experimental match and the lowest energy structure. 
For a given compound, rank=1 and $\Delta E=0$ when the experimental match is correctly calculated as the lowest energy structure. 
If there are multiple experimental matches for a compound, we take the minimum rank/energy. 
This occurs for cases where there are similar relaxed structures that match the same experimental structure, or there are multiple experimental structures in the case of polymorphic compounds. 
\\

Relaxations are performed with an initial stage of FIRE optimization \cite{fire} with a convergence threshold of 0.1 eV/Å, followed by BFGS optimization \cite{broyden1970convergence,fletcher1970new,goldfarb1970family,shanno1970conditioning} with a convergence threshold of 0.01 eV/Å.
The initial FIRE optimization helps improve the robustness of the relaxation for cases where initial forces are large. 
A very small fraction of the relaxations ($<0.01\%$) involved changes in bond connectivity, and were excluded from the ranking. 
The top 50 structures ranked by energy for each compound were then reranked using the harmonic approximation to the free energy at 300 K. 
Supercells were chosen by maximising the minimum lattice vector, subject to keeping the number of atoms in the supercell below 1182. 
This minimises error from the choice of supercell while keeping the GPU memory usage below OOM for all models. 
Further details of the experimental setup are provided in appendix \ref{app:az-eval-set}. 
\\

\autoref{fig:az_landscape} and \autoref{tab:az_model_metrics} show that energy reranking with CSP-MACE-Å performs comparably to the PBE-NP DFT, while including the harmonic approximation to the free energy significantly improves the accuracy to give the best overall performance. 
CSP-MACE-Å performs consistently across the full set, ranking the experimental match within the top 10 structures, and within 2 kJ/mol of the minimum energy across all compounds when using the free energy. 
MACE-POLAR-1 and UMA-OMC show worse performance, with the latter significantly worse. 
The improvement of CSP-MACE-Å over MACE-POLAR-1 shows the benefits of the intermolecular delta model trained on crystal structures, as well as the inclusion of long-range dispersion. 
\\

\begin{table}[h]
\centering
\caption{AZ evaluation set metrics averaged over the 19 compounds. We show the energy (E), and Helmholtz free energy (A) under the harmonic approximation at 300 K. $\Delta E$ is defined as the difference in energy between the experimental match and the lowest energy structure.}
\label{tab:az_model_metrics}
\begin{tabular}{lcc}
\hline
\textbf{Model} & \textbf{~Rank~~~} &  $\boldsymbol{\Delta E}$ \textbf{(kJ/mol)} \\
\hline
PBE-NP E   &  3.68 & 0.68 \\
CSP-MACE-Å E & 3.58   & 0.66 \\
\textbf{CSP-MACE-Å A}  & \textbf{2.11} & \textbf{0.36} \\
UMA-OMC E  & 25.26 & 1.75 \\
UMA-OMC A  & 23.26 & 1.59 \\
MACE-POLAR-1 E  & 5.58 & 1.32 \\
MACE-POLAR-1 A  & 3.00 & 0.58 \\
\hline
\end{tabular}
\end{table}

\subsection{Blind Test Set}

\begin{figure*}[!t]
    \centering
    \includegraphics[width=1.0\linewidth]{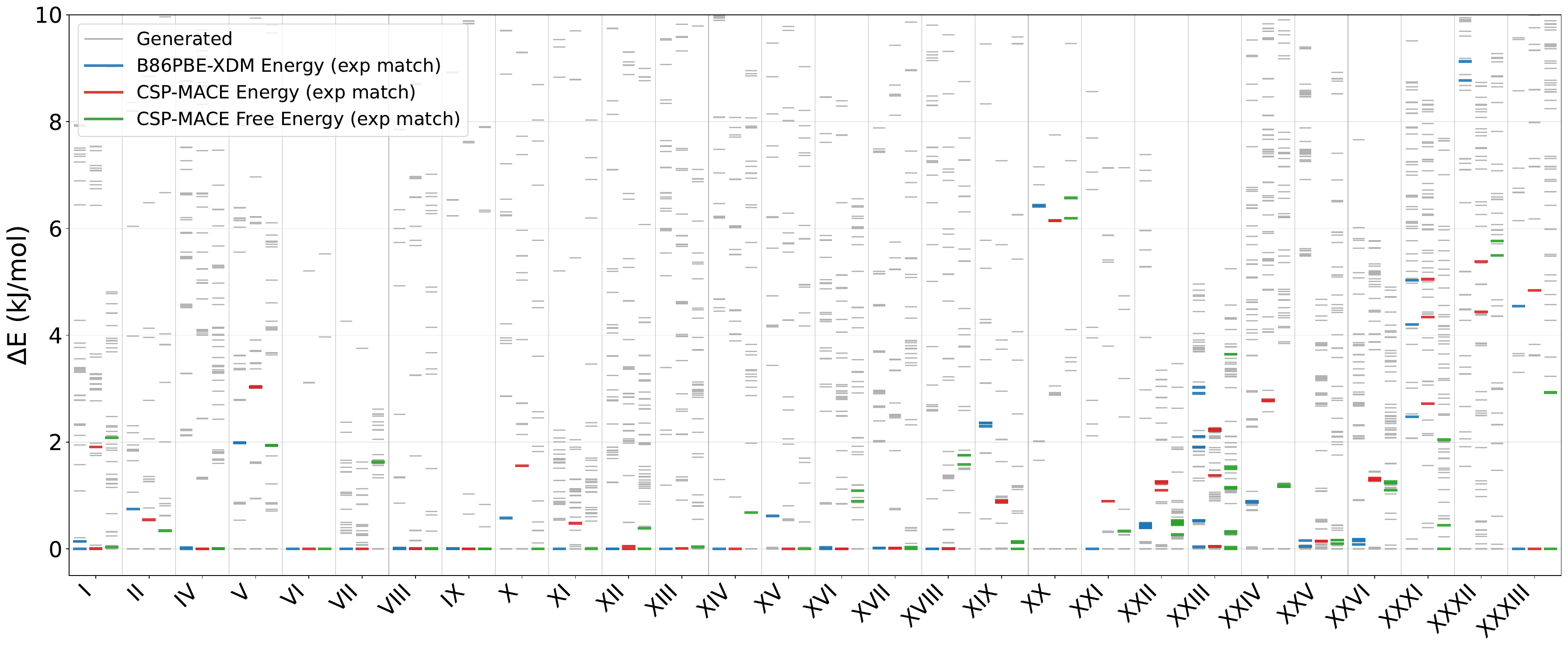}
    \caption{Blind test evaluation set relative energy landscape. The left hand column for each compound is B86bPBE-XDM DFT Energy ranking and the central column CSP-MACE-Å energy ranking, and the right hand column is CSP-MACE-Å free energy ranking under the harmonic approximation at 300 K. The blind test set includes both the generated structures submitted by participating groups and the experimental structures. For each compound, generated structures that do not match experiment are shown in grey; experimental structures and generated structures which match experimental structures are shown in colour. 
    }
    \label{fig:bt_landscape}
\end{figure*}

Our second evaluation set is composed of 28 compounds, including cocrystals and salts, from the seven CSP blind tests \cite{lommerse2000test,motherwell2002crystal,day2005third,day2009significant,bardwell2011towards,reilly2016report,hunnisett2024seventh_rank,hunnisett2024seventh_gen}. 
We use the structures collated in \citet{nickerson2025assessment} which includes both the submitted and experimental structures from each of the blind tests\footnote{We note that this set omits a small number of submitted structures from the previous blind tests.}. 
The structures from \citet{nickerson2025assessment} have been relaxed with B86bPBE-XDM DFT. 
Unlike the AZ Set in \autoref{sec:az_eval_set}, this set does not correspond to a standard end-to-end CSP workflow, since it combines submissions from multiple blind tests and uses structures already relaxed with B86bPBE-XDM.
Nevertheless, it still provides a useful benchmark for assessment. 
\\

Following \citet{nickerson2025assessment}, we remove duplicates before performing ranking, where duplicates are defined as having a powder difference score of less than 0.01, computed with critic2 \cite{otero2014critic2}. 
We then report the same ranking and $\Delta E$ metrics as \autoref{sec:az_eval_set}. 
The relaxation and free energy protocol for all models follows \autoref{sec:az_eval_set}. 
\\

\autoref{fig:bt_landscape} and \autoref{tab:bt_metrics} show that CSP-MACE-Å outperforms the alternative MLIP baselines, especially with reranking by the free energy under the harmonic approximation. 
When ranking by energy only, the performance of CSP-MACE-Å is marginally worse than B86bPBE-XDM DFT. 
Reranking with the harmonic free energy with CSP-MACE-Å improves the ranking quality to  surpass the performance of the B86bPBE-XDM DFT energy ranking. 
The improvement between the energy and free energy ranking is especially pronounced for compound XXXI, which is consistent with prior results for B86bPBE-XDM DFT reported by \citet{mayo2024assessment}.

\begin{table}[h]
\centering
\caption{Blind test set evaluation metrics averaged over the 28 compounds. We show the energy (E), and Helmholtz free energy (A) under the harmonic approximation at 300 K. $\Delta E$ is defined as the difference in energy between the experimental match and the lowest energy structure.}
\label{tab:bt_metrics}
\begin{tabular}{lcc}
\hline
\textbf{Model} & \textbf{~Rank~~~} &  $\boldsymbol{\Delta E}$ \textbf{(kJ/mol)} \\
\hline
B86bPBE-XDM E & 3.25 & 0.91 \\
CSP-MACE-Å E  & 3.86 & 0.94 \\
\textbf{CSP-MACE-Å A}   & \textbf{2.96} & \textbf{0.80} \\
UMA-OMC E   & 3.89 & 1.11 \\
UMA-OMC A   & 3.50 & 0.89 \\
MACE-POLAR-1 E  & 7.46 & 1.85 \\
MACE-POLAR-1 A  & 5.43 & 1.18 \\
\hline
\end{tabular}
\end{table}

\subsection{ROY}
\begin{figure*}[t]
    \centering
    \includegraphics[width=\linewidth]{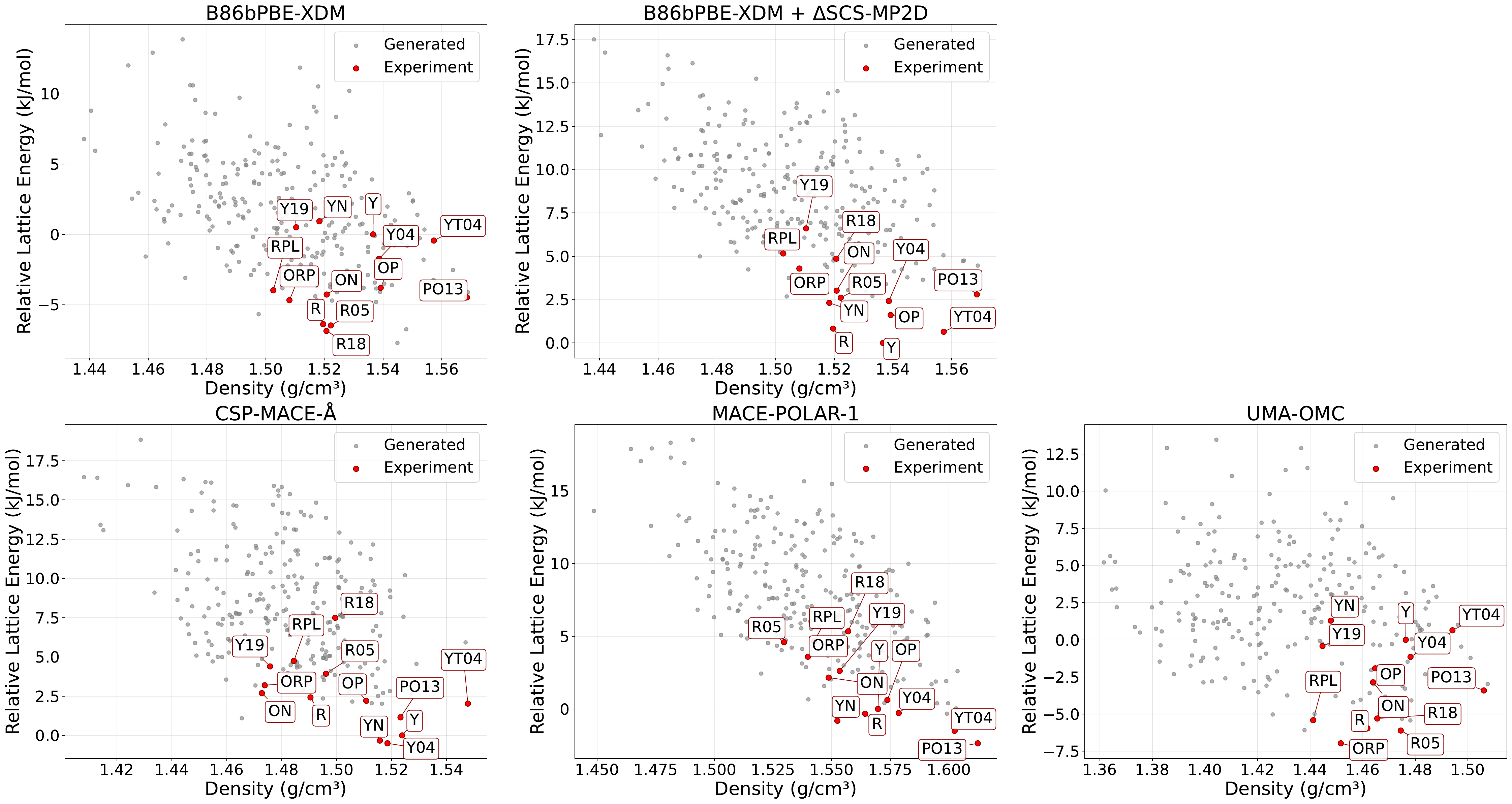}
    \caption{ROY energy landscape plots. Lattice energies are relative to Form Y which is the experimentally most stable polymorph. }
    \label{fig:roy}
\end{figure*}

Next, we evaluate CSP-MACE-Å on ROY (\textit{Red Orange Yellow}), which has 14 known polymorphs, holding the record for the organic crystal with the most characterized polymorphs \cite{weatherston2025polymorphic}.
Common DFT-D methods are inaccurate for CSP applied to ROY due to errors in its modelling of intramolecular interactions \cite{nyman2019accuracy,beran2022many}. 
Specifically, delocalization error causes these methods to predict the experimentally most stable yellow polymorph, Form Y, to lie several kJ/mol above orange and red polymorphs and above some generated structures not observed experimentally.
\citet{beran2022many} show that B86bPBE-XDM is inaccurate for ROY due to delocalisation error but this can be corrected by replacing the modelling of intramolecular interactions with spin-component-scaled dispersion-corrected
second-order Møller–Plesset perturbation theory (SCS-MP2D) with an equivalent form to \autoref{eqn:inter_intra}.
SCS-MP2D is a suitable choice because it has been shown to achieve near coupled cluster accuracy for conformational energies of isolated ROY molecules. 
\\

CSP-MACE-Å approximates B86bPBE-XDM DFT only for intermolecular interactions, while approximating the higher level of theory ωB97M-V/def2-TZVPD DFT for intramolecular interactions.
ROY therefore is a useful case study as CSP-MACE-Å should not suffer from the same poor performance as B86bPBE-XDM. 
Additionally, we evaluate the performance of MACE-POLAR-1 and UMA-OMC. 
We relax each structure from \citet{beran2022many} with the same settings as \autoref{sec:az_eval_set} and then analyze the energy landscape. 
\\

\autoref{fig:roy} shows that CSP-MACE-Å predicts the most stable Form Y polymorph within 0.5 kJ/mol of the minimum; an error of similar magnitude to the average accuracy on the AZ set and the blind test set. 
On the other hand, B86bPBE-XDM places Form Y more than 5 kJ/mol above the minimum. 
Using the SCS-MP2D intramolecular correction to B86bPBE-XDM DFT is shown to correctly rank Form Y as the most stable polymorph. 
UMA-OMC performs similarly poorly to B86bPBE-XDM, likely due to also suffering from delocalisation error. 
MACE-POLAR-1 ranks Form Y within 2.5 kJ/mol of the minimum, outperforming UMA-OMC and B86bPBE-XDM due to superior intramolecular accuracy.
However, it performs worse than CSP-MACE-Å and B86bPBE-XDM+$\Delta$SCS-MP2D indicating intermolecular accuracy is limiting. 
Thus, ROY provides a useful case study illustrating the contribution of both the intramolecular and intermolecular models to the accuracy of CSP-MACE-Å.

\subsection{Thermodynamic Stability of Different Polymorphs}
For our final evaluation, we collate five compounds from previous AZ studies of polymorphic thermodynamic stability \cite{broo2016transferable,andrews2021derisking,wu2025discovery,sun2021current,putra2025mechanistic} to evaluate CSP-MACE-Å's capacity to accurately model temperature-dependent polymorphic stability. 
\\

For each compound, we relax each experimental polymorph using the same protocol as \autoref{sec:az_eval_set} but with a relaxation criterion of 0.001 eV/Å. 
Following this, we calculate the free energy under the harmonic approximation at intervals of 20 K from 0 to 600 K.
Supercells were chosen by maximising the minimum lattice vector, subject to keeping the number of atoms in the supercell below 2048.
The larger supercells and stricter relaxation criterion are conservatively chosen as the free energy differences between polymorphs at different temperatures are often very subtle. 
\\

Three of the compounds (Mexiletine Hydrochloride, the Triazolo-pyrimidine, AZD5462) have polymorphs with disordered sites with partial occupancy. 
For these disordered structures the free energy is separately calculated for each combination of states of the disordered sites.
These state-specific free energies, together with the occupancy fractions, are then combined to give an overall free energy of the disordered polymorph using the ideal mixing approximation: 
\begin{equation}
G(T) = \sum_{i=1}^{N} p_i \, G_i(T) + k_B T \sum_{i=1}^{N} p_i \ln p_i
\end{equation}
where $G_i(T)$ is the Gibbs free energy of state $i$ at temperature $T$, $N$ is the number of distinct states, $p_i$ is the fractional occupancy of state $i$, and $k_B$ is the Boltzmann constant.
The second term, $k_B T \sum_{i=1}^{N} p_i \ln p_i$, is the ideal configurational entropy contribution to the free energy.


\subsubsection{Sulfathiazole}
Sulfathiazole has five known polymorphs, which exhibit enantiotropic polymorphism. 
Experimental data suggests that at low/room temperatures the relative stability of the polymorphs is IV $\approx$ III $>$ II $>$ I $>$ V, 
while between 283 K and 332 K the relative stability is III $>$ II $>$ IV $>$ V $>$ I ,
and at high temperatures (450 K) the relative stability is I $>$ V $>$ IV $>$ III $>$ II \cite{parmar2007polymorph,munroe2012relative,sovago2014experimental}. 
\\

\begin{figure}[h]
    \centering
    \includegraphics[width=\linewidth]{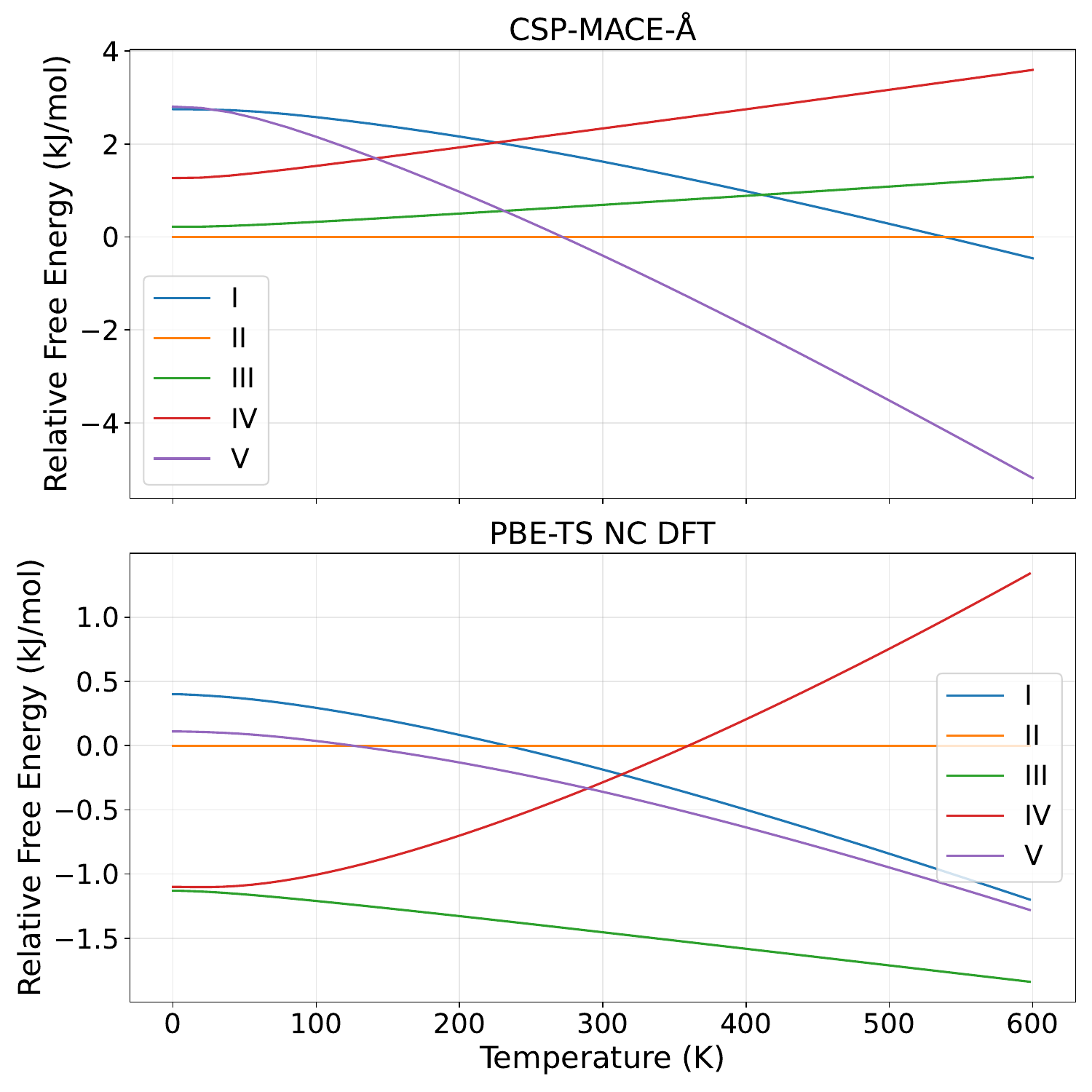}
    \caption{Sulfathiazole thermodynamic stability predictions with CSP-MACE-Å and PBE-TS NC DFT using the harmonic approximation to the free energy.}
    \label{fig:suthaz_temp}
\end{figure}

\autoref{fig:suthaz_temp} show the free energy with temperature relationships for CSP-MACE-Å, as well as with PBE-TS NC DFT which was studied by \citet{broo2016transferable}. 
CSP-MACE-Å correctly predicts the broad trend: the relative stability of V and I increase with temperature, whereas the remaining compounds relative stability decrease with temperature. 
However, CSP-MACE-Å does not recover the exact polymorph ordering at either low or high temperature, with II incorrectly appearing to be the most stable at low temperature (instead of IV and III), and V appearing to be the most stable at high temperatures (instead of I) when in fact it is the second most stable. 
PBE-TS NC DFT has qualitatively similar behavior: the broad trends were captured, but the exact ordering of polymorph stability is not reproduced across temperatures.

\subsubsection{Mexiletine Hydrochloride}

Mexiletine hydrochloride is a salt with five solid-form types. 
Form I, II and III are anhydrous, and Types A and B are families of channel solvates with the addition of a nonsolvated form isostructural to the Type A solvate \cite{andrews2021derisking}.
Form II has disorder with two distinct conformations of the mexiletine molecule. 
Here we study the relative stability of the anhydrous forms: Form I, II, III and the nonsolvated Type A form.
Experimental data suggests that the most stable polymorphs are Form I at room temperature, Form III between 421 K and 440 K, and Form II at high temperature \cite{andrews2021derisking}.
\\

\begin{figure}[h]
    \centering
    \includegraphics[width=\linewidth]{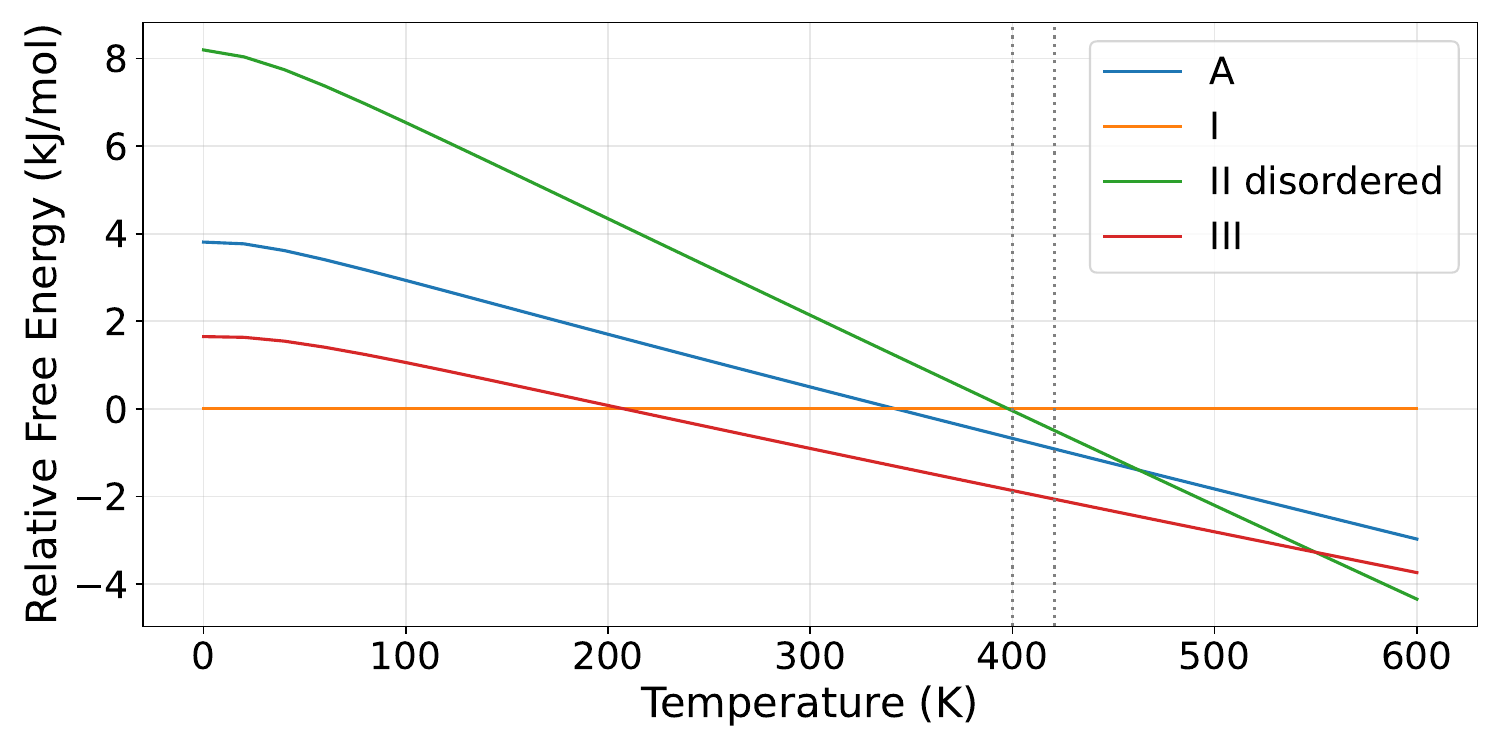}
    \caption{Mexiletine hydrochloride thermodynamic stability predictions with CSP-MACE-Å using the harmonic approximation to the free energy. Experimental transition temperatures of I $\rightarrow$ II $\rightarrow$ III indicated with vertical dotted lines.}
    \label{fig:mexi_temp}
\end{figure}

\autoref{fig:mexi_temp} shows that CSP-MACE-Å correctly predicts Form I as the most stable at low temperature, Form II at high temperature, and Form III at an intermediate temperature. 
CSP-MACE-Å incorrectly predicts Form III to be more stable at a wider temperature range, from approximately 220 K to 550 K, rather than the true range of 421 K to 440 K. 

\subsubsection{AZD1305} 

\begin{figure}[h]
    \centering
    \includegraphics[width=\linewidth]{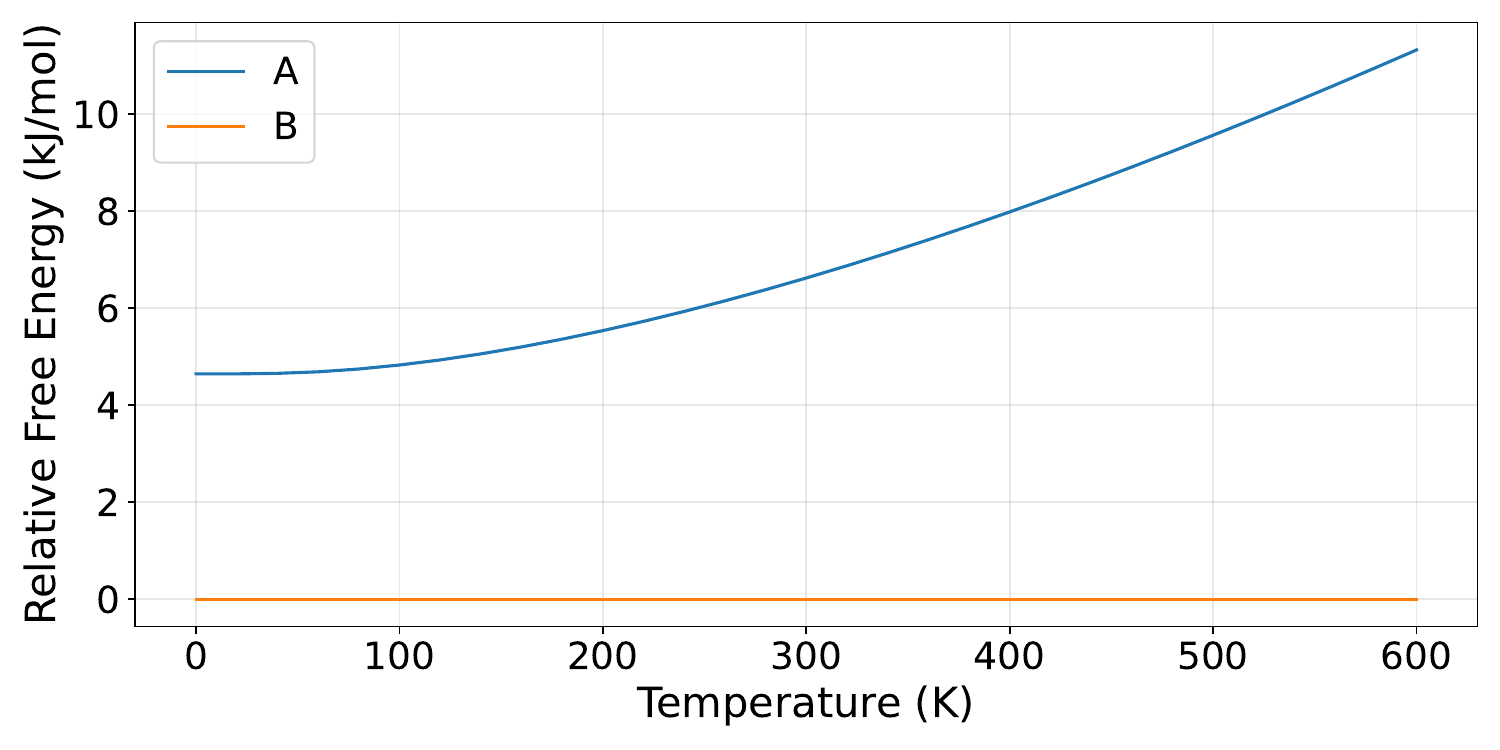}
    \caption{AZD1305 thermodynamic stability prediction with CSP-MACE-Å using the harmonic approximation to the free energy.}
    \label{fig:AZD1305_temp}
\end{figure}

AZD1305 is a crystalline oxabispidine pharmaceutical with two polymorphs, Form A and B. 
Experimental data suggests the forms have a monotropic relationship, with Form B more stable at all temperatures, while their relative stability becomes closer at elevated temperatures \cite{sigfridsson2012preformulation}. 
\autoref{fig:AZD1305_temp} shows that CSP-MACE-Å correctly predicts the monotropic relationship and is therefore consistent with the experimental data, although it fails to predict closer relative stability at higher temperatures. 
\citet{sun2021current} estimated the free energies using XtalPi which combines DFT-D calculations of the energy with an additional temperature dependent term based off forcefield calculations, but incorrectly predicted an enantiotropic relationship.  

\subsubsection{Triazolo-pyrimidine} 

\begin{figure}[h]
    \centering
    \includegraphics[width=\linewidth]{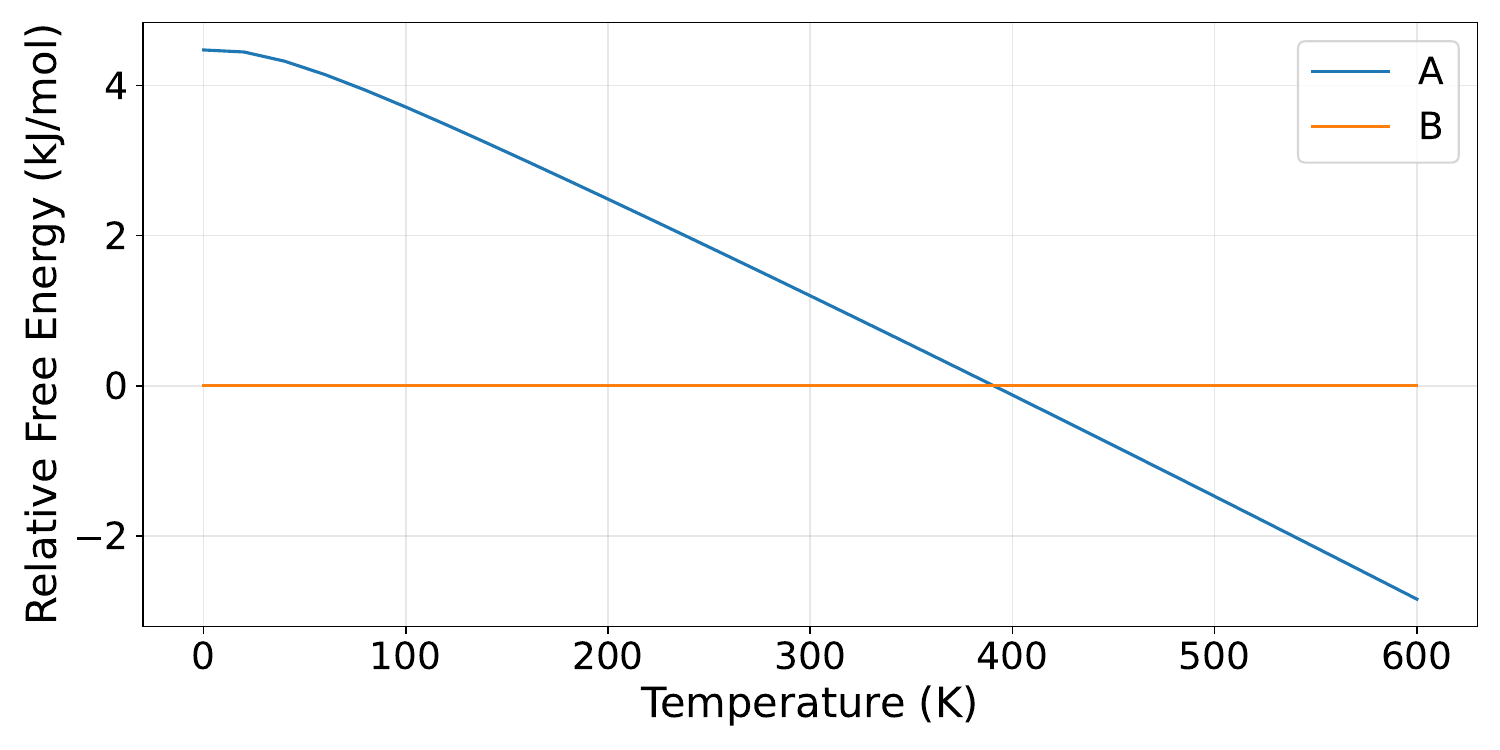}
    \caption{Triazolo-pyrimidine compound thermodynamic stability predictions with CSP-MACE-Å using the harmonic approximation to the free energy.}
    \label{fig:TUJMIY_temp}
\end{figure}

\citet{wu2025discovery} presents a study of a triazolo-pyrimidine compound which is an adenosine receptor antagonist with two stable forms.
Experimental data suggests Form B is more stable at room temperature and Form A is more stable at high temperature. \autoref{fig:TUJMIY_temp} shows that the predictions of CSP-MACE-Å are accurate for this compound, with free energies that are consistent with the experimental data. 

\subsubsection{AZD5462}
\begin{figure}[h]
    \centering
    \includegraphics[width=\linewidth]{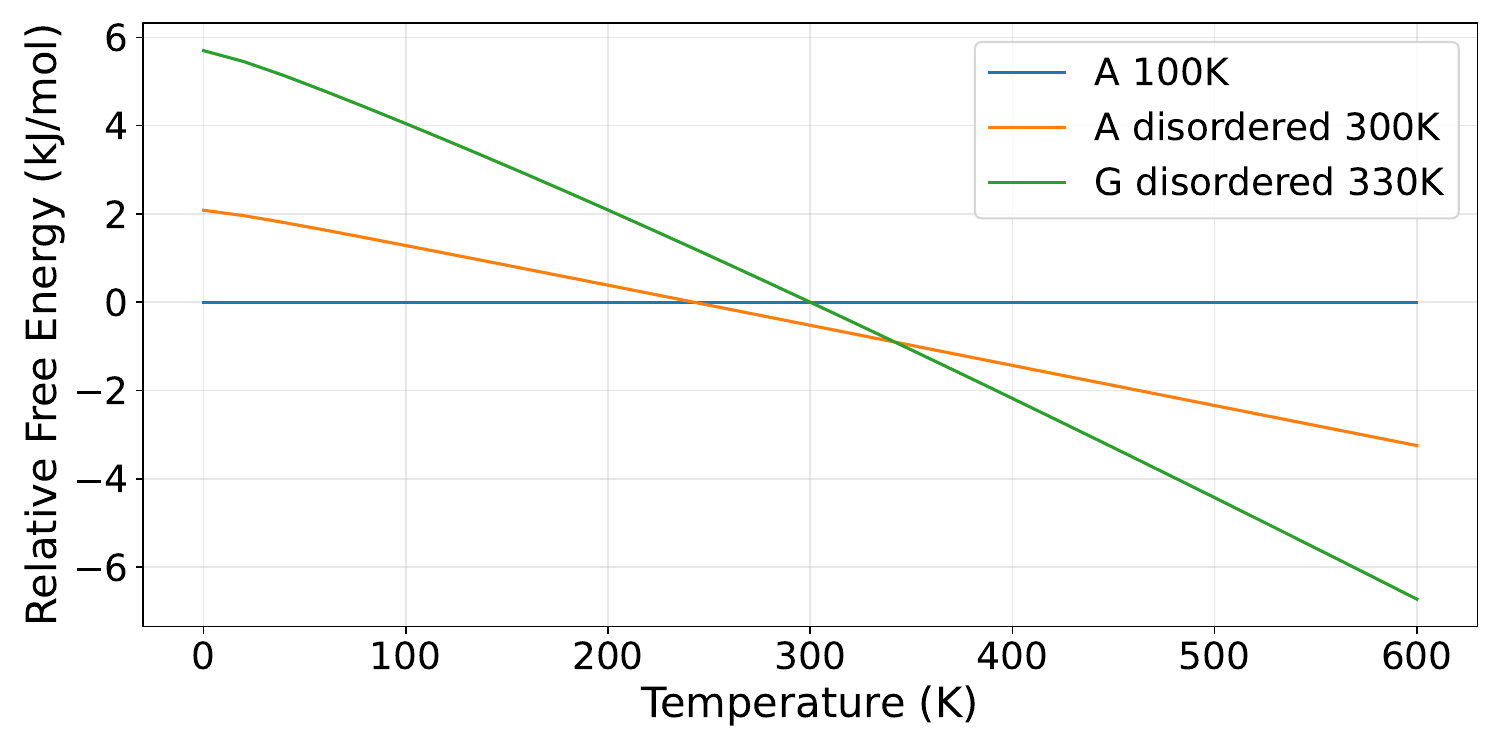}
    \caption{AZD5462 thermodynamic stability predictions with CSP-MACE-Å using the harmonic approximation to the free energy.}
    \label{fig:AZD5462_temp}
\end{figure}

AZD5462 is an agonist of the relaxin family peptide receptor, being developed for the treatment of cardiorenal
diseases \cite{putra2025mechanistic}.
It has two forms which exhibit enantiotropic polymorphism. 
Form G is disordered, with two sites each containing two distinct conformations, and is the most stable polymorph at 330 K. 
Form A has an ordered form which is the most stable at 100 K, and a disordered form with a single site containing two distinct conformations, which is thermodynamically more favourable than the ordered form at 300 K. 
\autoref{fig:AZD5462_temp} shows that CSP-MACE-Å's predictions closely match the experimental data, whereby the ordered form A is the most stable at low temperatures, the disordered form A at an intermediate temperature around 300 K, and form G at elevated temperatures. 

\section{Conclusion}
We have presented an evaluation of CSP-MACE-Å, an MLIP designed to replace DFT within CSP. 
CSP-MACE-Å builds on the MACE-POLAR architecture to include long-range dispersion and a delta model which learns to capture subtle intermolecular interactions of crystal structures. 
We evaluated CSP-MACE-Å on two benchmark sets: 19 compounds (including a salt, each with a large structure pool) collated from AstraZeneca's previous CSP publications, and 28 compounds (including cocrystals and salts) from the seven CSP blind tests. 
On both sets, CSP-MACE-Å outperforms existing MLIPs and provides accuracy comparable to DFT.
On five compounds, predicting the thermal free energy under the harmonic approximation with CSP-MACE-Å is shown to broadly match experimental data on trends in temperature dependent stability of different polymorphs, although the exact ordering of polymorphs across temperatures was not always correct. 
In this work we have kept the standard CSP workflow fixed, and replaced DFT with CSP-MACE-Å. 
Future work will explore how CSP workflows themselves can be adapted to better utilize the benefits that CSP-MACE-Å offers. 
Specifically, CSP-MACE-Å could be incorporated into structure generation rather than only the final ranking step. 

\section{Acknowledgments}
We thank Erin Johnson for help with the blind test evaluation.

\bibliography{refs}

\appendix
\section{Appendix}

\subsection{Experimental details and Further Results}
\subsubsection{AZ Evaluation Set}
\label{app:az-eval-set}

\begin{figure*}[!t]
    \centering
    \begin{minipage}[t]{0.48\textwidth}
        \centering
        \includegraphics[width=\linewidth]{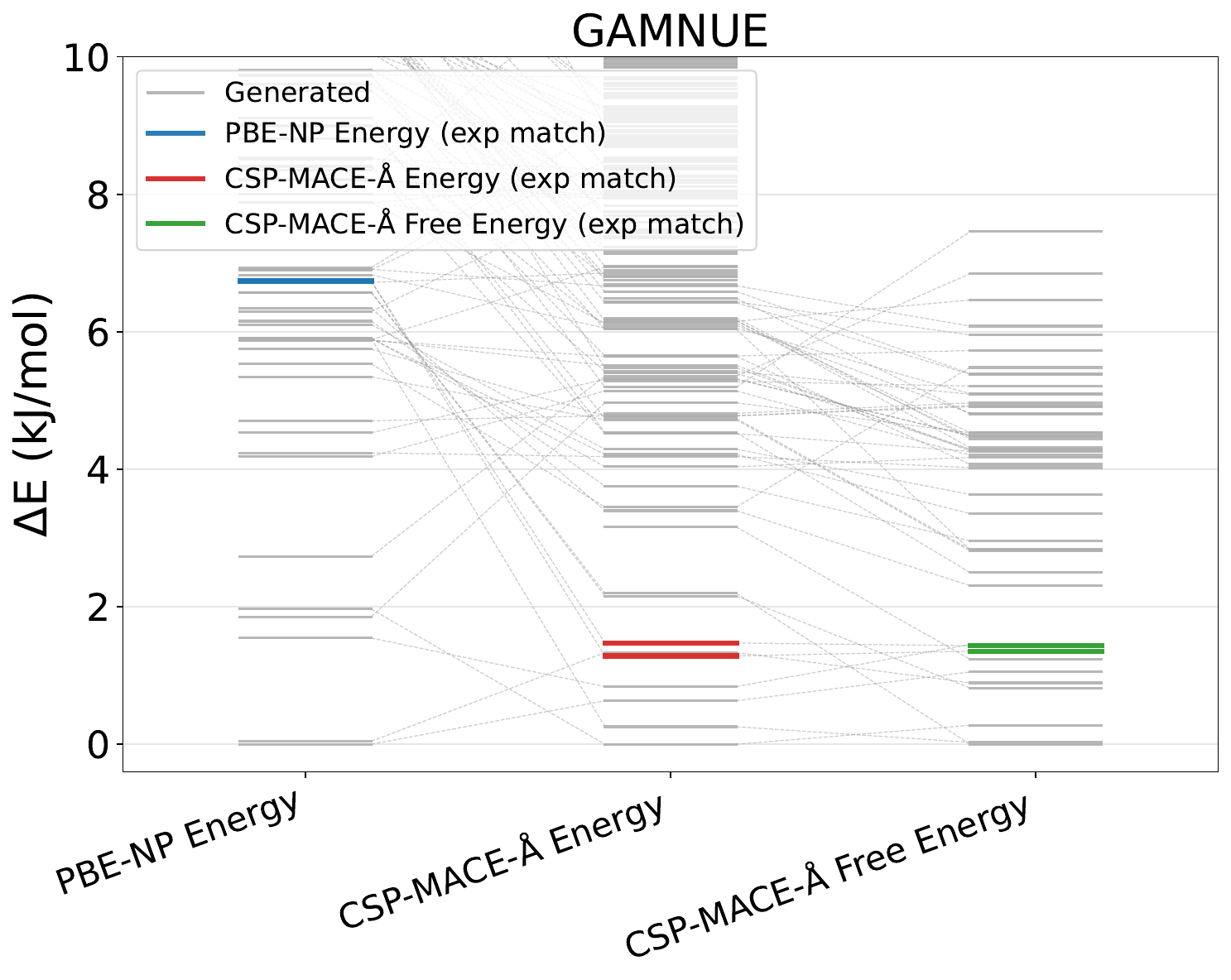}
        \caption{GAMNUE relative energy landscape. Connections are shown between same initial (pre-relaxation) structures across methods.}
        \label{fig:GAMNUE}
    \end{minipage}\hfill
    \begin{minipage}[t]{0.48\textwidth}
        \centering
        \includegraphics[width=\linewidth]{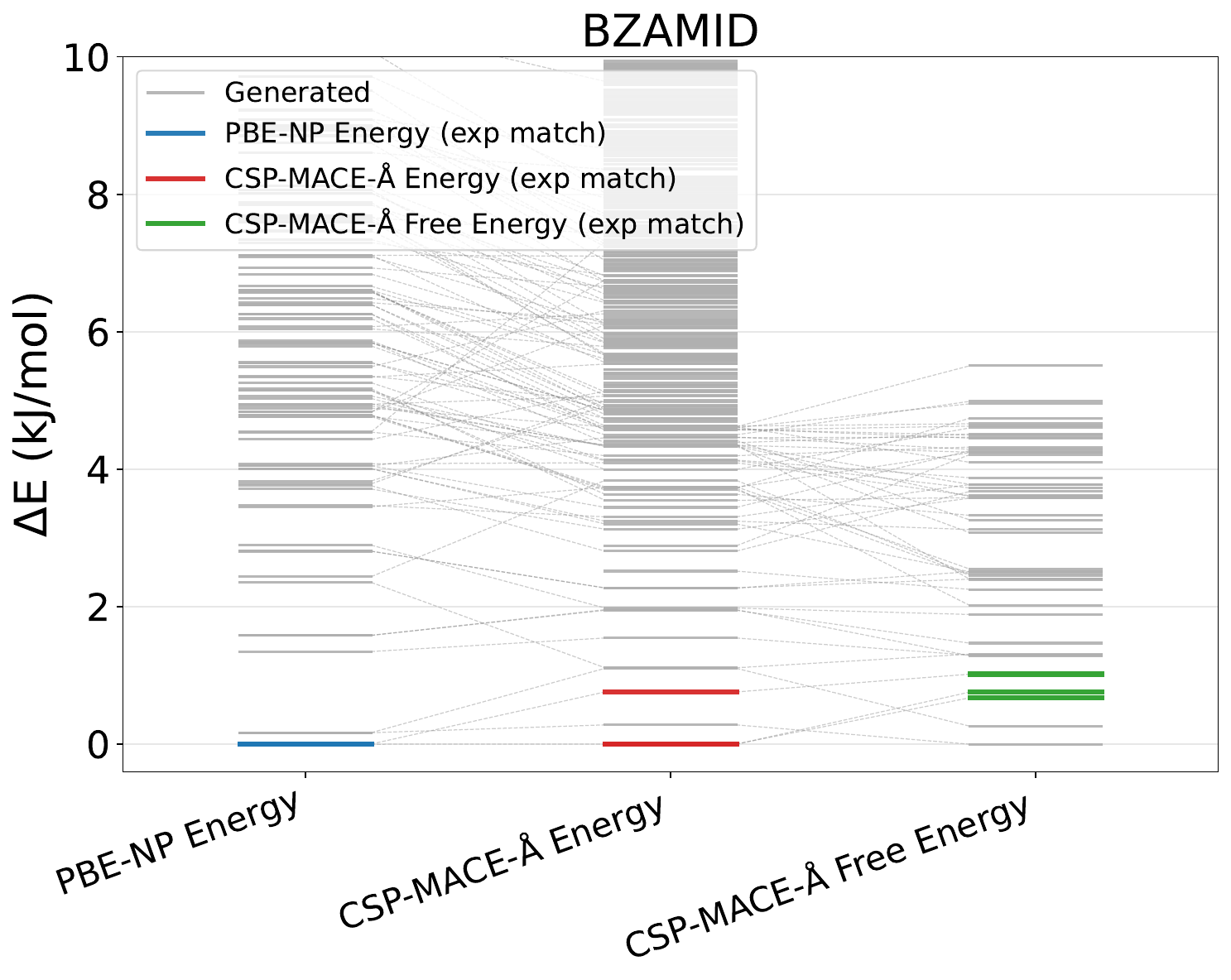}
        \caption{BZAMID relative energy landscape. Connections are shown between same initial (pre-relaxation) structures across methods.}
        \label{fig:BZAMID}
    \end{minipage}
\end{figure*}

\begin{figure*}[t]
    \centering
    \includegraphics[width=\linewidth]{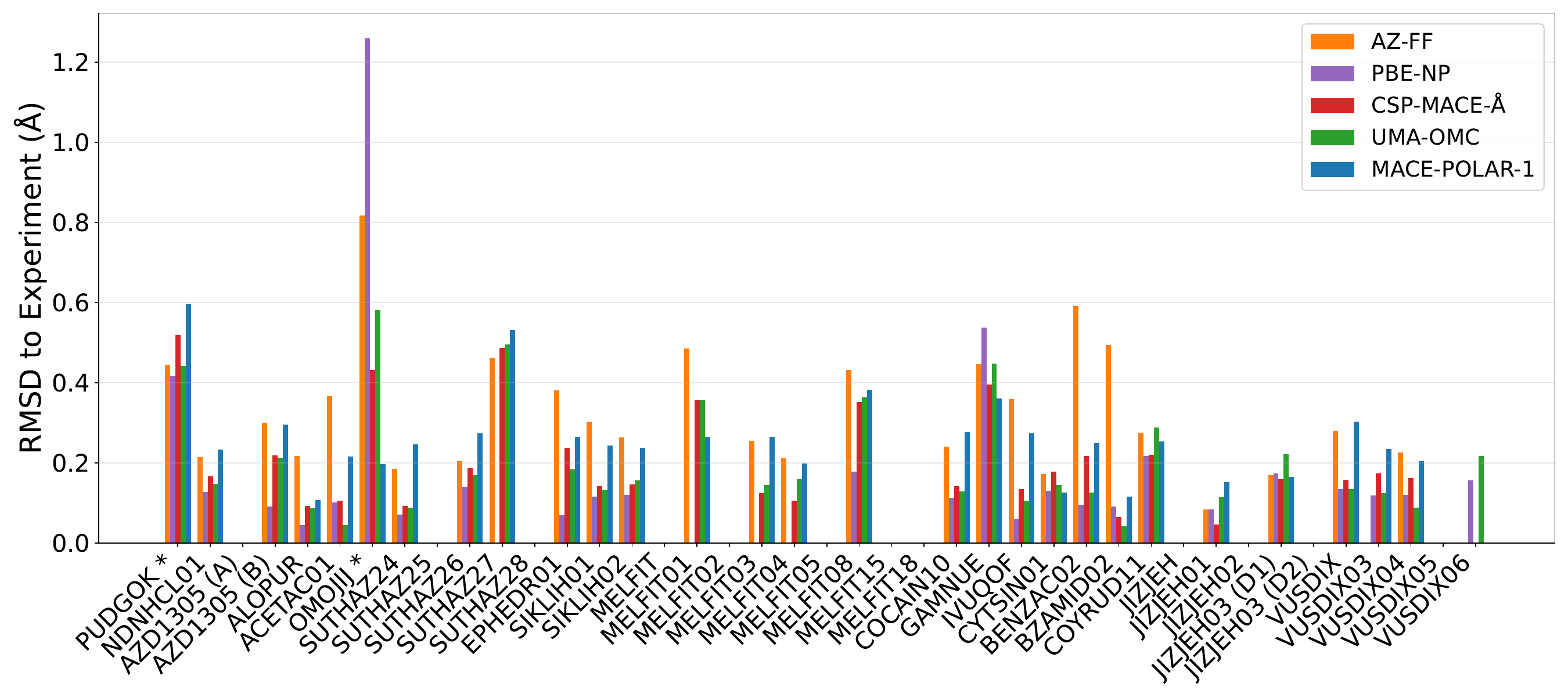}
    \caption{RMSD for matches of relaxed structures to experimental structures. AZD1305 experimental structures do not have CCDC codes - forms A and B are instead directly noted. JIZJEH03 contains a disordered and therefore split into two structures D1 and D2. PUDGOK contains a partial match for AZ-FF and PBE-NP with 9/15 and 3/15 molecules overlayed respectively. OMOJIJ contains a partial match for all methods with an 11/15 molecule overlay for AZ-FF and MACE-POLAR-1 and a 12/15 overlay for PBE-NP, CSP-MACE-Å and UMA-OMC. For structures where no bar is present, no matches (incl. partial matches) were found.}
    \label{fig:az_rmsd}
\end{figure*}

\begin{figure*}[t]
    \centering
    \includegraphics[width=\linewidth]{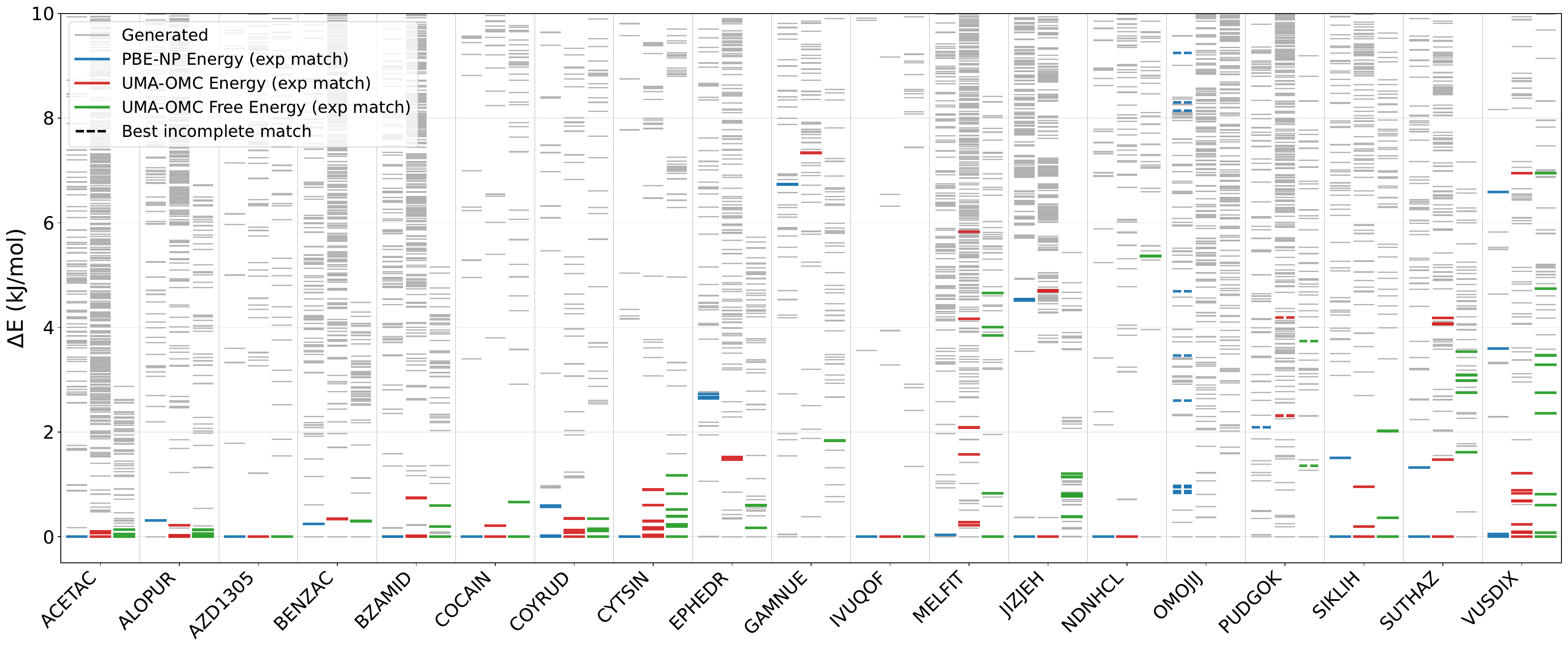}
    \caption{AZ evaluation set relative energy landscape. The left hand column for each compound is PBE-NP DFT Energy ranking and the central column UMA-OMC energy ranking, and the right hand column is UMA-OMC free energy ranking under the harmonic approximation at 300 K.}
    \label{fig:az_landscape_omc}
\end{figure*}

\begin{figure*}[t]
    \centering
    \includegraphics[width=\linewidth]{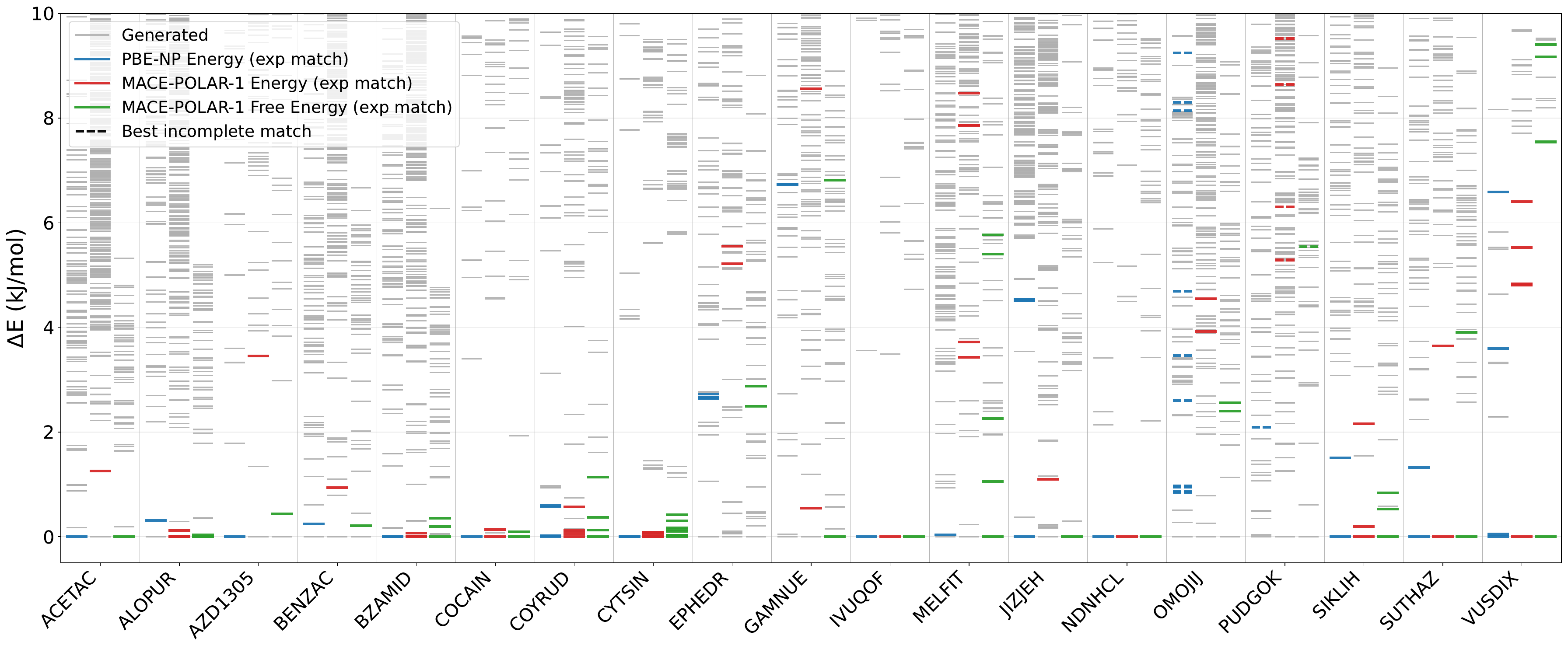}
    \caption{AZ evaluation set relative energy landscape. The left hand column for each compound is PBE-NP DFT Energy ranking and the central column MACE-POLAR-1 energy ranking, and the right hand column is MACE-POLAR-1 free energy ranking under the harmonic approximation at 300 K.}
    \label{fig:az_landscape_polar}
\end{figure*}

The number of structures from each stage of CSP on the AZ set is not consistent across compounds, as this set is collated from multiple CSP analyses. 
MLIPs reranking is applied to all available AZ-FF structures (typically around 1000 structures), while PBE-NP DFT relaxation is done on a smaller subset *around 100 structures).
\autoref{tab:az_structure_counts} describes the number of structures MLIP and DFT relaxation are applied to, for each compound.
\autoref{tab:az_ccdc_ids} provides the CCDC identifiers for all experimental structures.
For CSP-MACE-Å and MACE-POLAR-1 there is consistently an experimental match within the top 50 structures, and thus reranking via the free energy for only these top 50 structures is appropriate. 
However, for UMA-OMC applied to the OMOJIJ compound, the experimental match was not within the top 50 structures (it ranked 418), so the full set of 452 structures were reranked under the harmonic approximation.
For the PUDGOK compound, none of the models had complete matches satisfying with a setting of $RMSD_{15} \mathord{<} 0.6\text{ \AA}$ so partial matches were allowed. 
On PUDGOK, for PBE-NP DFT, CSP-MACE-Å, UMA-OMC a 12 molecule cluster overlay was successfully fit to the experimental structure, and for MACE-POLAR-1 a 11 molecule cluster overlay was found by the COMPACK algorithm.
For OMOJIJ PBE-NP DFT had a partial match for the experimental structure with a 3 molecular cluster overlay.
\autoref{fig:az_rmsd} provides RMSD values for matches between experimental structures and matched structures for each relaxation method.
\autoref{fig:GAMNUE} and \autoref{fig:BZAMID}  show individual compound visualisations of the energy landscape, the former where CSP-MACE-Å significantly outperforms the PBE-NP DFT, and the latter an example of a case where the harmonic free energy approximation worsens performance. 
Lastly, we provide the ranking landscape plots for the UMA-OMC and MACE-POLAR-1 models relative to the PBE-NP reference in \autoref{fig:az_landscape_omc} and \autoref{fig:az_landscape_polar}. 
\\

\begin{table}[h]
\centering
\caption{Number of MLIP and DFT relaxed structures per compound. }
\label{tab:az_structure_counts}
\begin{tabular}{lrr}
\toprule
Compound & MLIP & DFT \\
\midrule
PUDGOK      &  1{,}000 &   99 \\
NDNHCL      &    894 &   82 \\
AZD1305     &  1{,}000 &  100 \\
ALOPUR      &  1{,}000 &  100 \\
ACETAC      &    893 &  100 \\
OMOJIJ      &    452 &  100 \\
SUTHAZ      &  1{,}000 &   99 \\
EPHEDR      &  1{,}000 &  100 \\
SIKLIH      &  1{,}000 &   79 \\
MELFIT &  2{,}212 &  395 \\
COCAIN      &    733 &  100 \\
GAMNUE      &    640 &   99 \\
IVUQOF      &  1{,}000 &  100 \\
CYTSIN      &  1{,}000 &  100 \\
BENZAC      &  1{,}000 &  100 \\
BZAMID      &  1{,}000 &  100 \\
COYRUD      &  1{,}000 &  100 \\
Mexilitine  &  1{,}000 & 1{,}000 \\
VUSDIX &  5{,}711 &  101 \\
\hline
\end{tabular}
\end{table}

\begin{table}[h] 
\centering 
\caption{CCDC codes for experimental structures.} 
\label{tab:az_ccdc_ids} 
\begin{tabular}{lr} \toprule
Compound & CCDC Codes \\
\midrule
ACETAC  & ACETAC01 \\
ALOPUR  & ALOPUR \\
AZD1305 & N/A; Forms A and B from \cite{sigfridsson2012preformulation} \\
BENZAC  & BENZAC02 \\
BZAMID  & BZAMID02 \\
COCAIN  & COCAIN10 \\
COYRUD  & COYRUD11 \\
CYTSIN  & CYTSIN01 \\
EPHEDR  & EPHEDR01 \\
GAMNUE  & GAMNUE \\
IVUQOF  & IVUQOF \\
JIZJEH  & JIZJEH, JIZJEH\{01--03\} \\
MELFIT  & MELFIT, MELFIT\{01--05,08,15,18\} \\
NDNHCL  & NDNHCL01 \\
OMOJIJ  & OMOJIJ \\
PUDGOK  & PUDGOK \\
SIKLIH  & SIKLIH01, SIKLIH02 \\
SUTHAZ  & SUTHAZ\{24--28\} \\
VUSDIX  & VUSDIX, VUSDIX\{01,03,04,06,07\} \\
\bottomrule
\end{tabular}
\end{table}



\newpage
\subsubsection{Blind Test Evaluation Set}
\label{app:az-eval-set}
MACE-POLAR-1 did not have an experimental match ranked within the top 50 for compound XXXII (it ranked 93) and therefore the top 150 structures were reranked under the harmonic approximation. 
\autoref{fig:az_landscape_omc} and \autoref{fig:az_landscape_polar} show the ranking landscape plots for the UMA-OMC and MACE-POLAR-1 models relative to the B86PBE-XDM DFT reference. 

\begin{figure*}[t]
    \centering
    \includegraphics[width=1.0\linewidth]{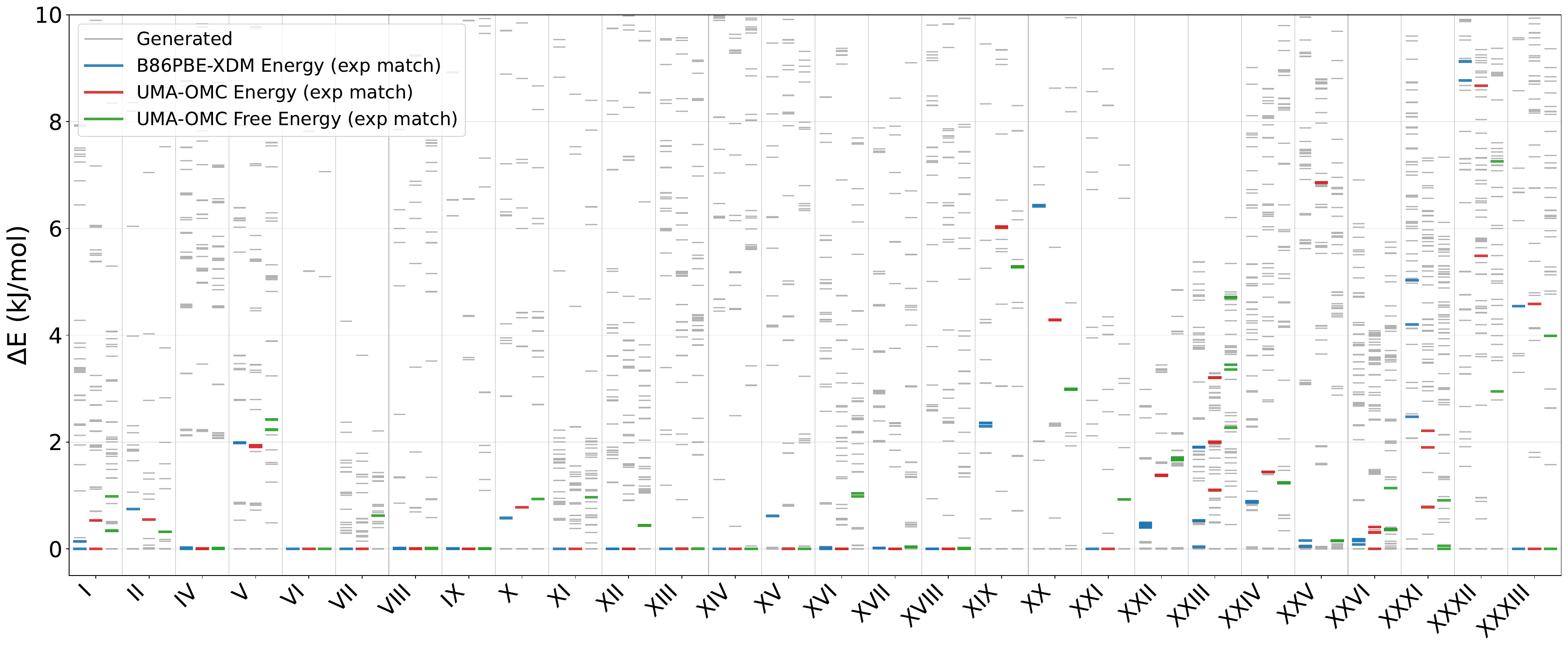}
    \caption{Blind test evaluation set relative energy landscape. The left hand column for each compound is B86bPBE-XDM DFT Energy ranking and the central column UMA-OMC energy ranking, and the right hand column is UMA-OMC free energy ranking under the harmonic approximation at 300 K.
    }
    \label{fig:bt_landscape_omc64}
\end{figure*}

\begin{figure*}[t]
    \centering
    \includegraphics[width=1.0\linewidth]{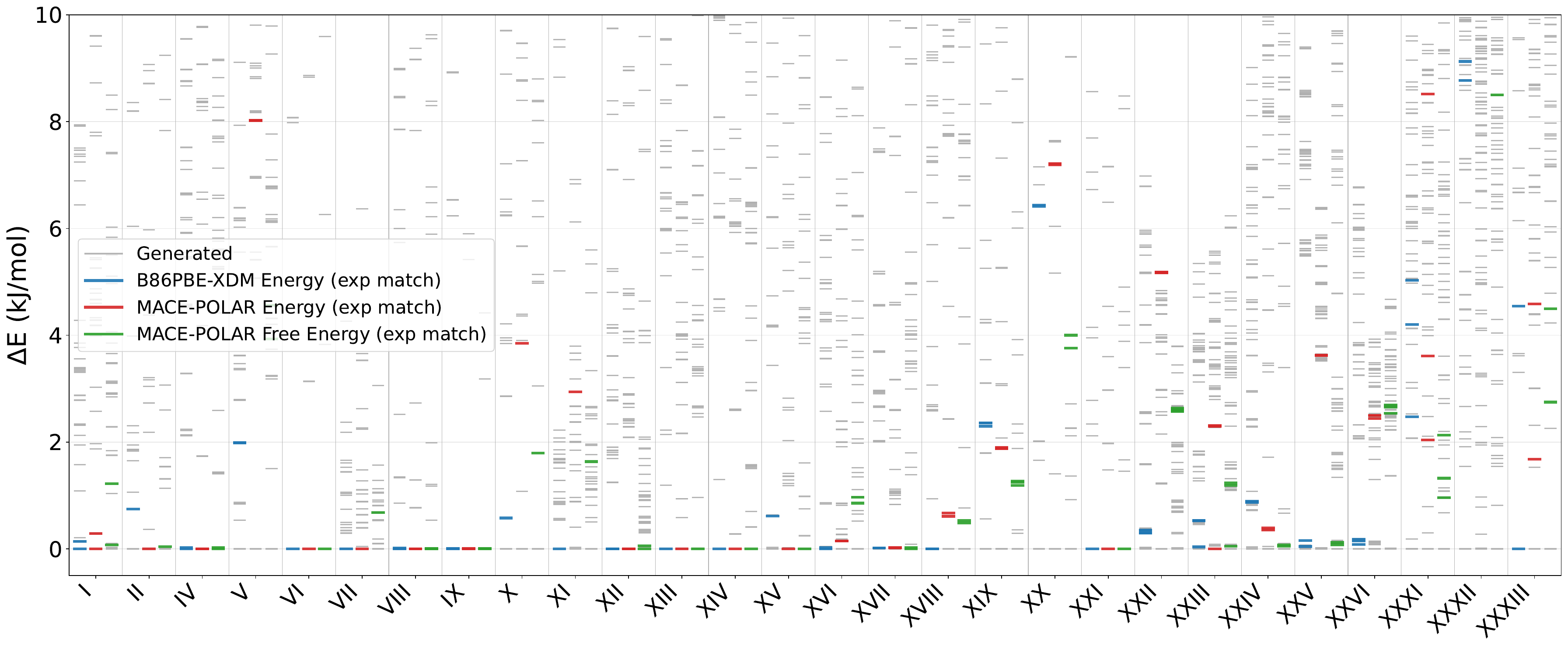}
    \caption{Blind test evaluation set relative energy landscape. The left hand column for each compound is B86bPBE-XDM DFT Energy ranking and the central column  MACE-POLAR-1 energy ranking, and the right hand column is  MACE-POLAR-1 free energy ranking under the harmonic approximation at 300 K.
    }
    \label{fig:bt_landscape_polar}
\end{figure*}

\end{document}